# Magnetization reversal process and nonlinear magneto-impedance in Cu/NiFe and Nb/NiFe composite wires

A.S. Antonov [a], N.A. Buznikov [a,*], A.B. Granovsky [b], I.T. Iakubov [a],
A.F. Prokoshin [c], A.L. Rakhmanov [a], A.M. Yakunin [a]

[a] *Institute for Theoretical and Applied Electrodynamics, Russian Academy of Sciences, 125412 Moscow, Russia*

[b] *Department of Physics, M.V. Lomonosov Moscow State University, 119992 Moscow, Russia*

[c] *Central Research Institute of Ferrous Metallurgy, 107005 Moscow, Russia*

**Abstract**

The magnetization reversal of Cu/NiFe and Nb/NiFe composite wires carrying AC current is studied. The frequency spectrum of a voltage induced in a pick-up coil wound around the wire is analyzed. The frequency spectrum is shown to consist of even harmonics within a wide range of AC current amplitudes and longitudinal DC magnetic fields. The strong dependencies of the harmonic amplitudes on the DC field are found. The results obtained may be of importance for the design of weak magnetic field sensors.

---

* Corresponding author. Fax: + 7-095-484-26-33.

*E-mail address:* n_buznikov@mail.ru (N.A. Buznikov)

# 1. Introduction

The giant magneto-impedance (GMI) in magnetic materials has been studied intensively in the last few years. The growing interest in GMI is supported by possible use of the effect in various technological applications, in particular, for the design of weak magnetic field sensors. Recently, GMI has been observed in electroplated composite wires consisting of a highly conductive inner core and an outer soft magnetic shell [1–5]. The presence of the highly conductive inner core in the composite wires may result in significant increase of the amplitude of GMI effect [6]. GMI can be observed at sufficiently low AC current amplitudes, when the output voltage is proportional to the sample impedance. At higher AC current amplitudes the magnetization reversal of the sample takes place. This process is accompanied by the appearance of a nonlinear voltage response at the sample ends and in the pick-up coil wound around the sample [1,7,8]. Due to the strong magnetic field dependence, this nonlinear magneto-impedance effect is also promising for possible technological applications.

The aim of the present paper is to study the nonlinear magneto-impedance in Cu/NiFe and Nb/NiFe composite wires fabricated by a modified cold-drawn technique [9]. The frequency spectrum of the voltage induced in the pick-up coil wound around the wire is analyzed as a function of AC current amplitude and longitudinal DC magnetic field. It is shown that even harmonics dominate in the frequency spectrum. A high sensitivity of the even harmonic amplitudes to the DC magnetic field is demonstrated. The experimental results are interpreted in terms of the Faraday law and the quasi-static Stoner–Wohlfarth model [10].

# 2. Experimental

The Cu/NiFe and Nb/NiFe composite wires were produced by pressing the composite rod blank, its continuos drawing and subsequent annealing [9]. The blank consisted of Cu or Nb rod with diameter of $0.67\,cm$ pressed in $Ni_{82}Fe_{18}$ tube with the outer diameter of $1\,cm$. The rod blank was drawn through the diamond dies and the obtained composite wires were annealed at $750^oC$ for a period of $1\,h$ in a vacuum of $10^{-3}\,Pa$. The diameter $d$ of the inner nonmagnetic core was in $40-100\,\mu m$ range, and the permalloy outer shell was $10-25\,\mu m$ thick. A ratio of $D/d$ (where $D$ is the wire diameter) was about 1.5 for all samples. The length of the studied samples was 1 cm. The anisotropy field $H_a$ of the wires lied in the range $3-8$ Oe, and the saturation magnetization $M$ was about 800 G [9].

A detailed description of the experimental set-up used can be found in Ref. [8]. The amplitude $I_0$ of the AC current flowing through the wire was changed from 0 to 200 mA, and the current frequency $f$ was varied from 0.1 to 1 MHz. The longitudinal DC magnetic field $H_e$ was created by a solenoid oriented along the wire axis, and was varied from $-50$ to 50 Oe. The pick-up coil was wound around the wire. The number of turns $N$ in the pick-up coil was 40. In the experiments the harmonic amplitudes of the voltage $V$ at the leads of the pick-up coil were measured by HP4395A spectrum analyzer as a function of the AC current amplitude and the longitudinal DC magnetic field.

**3. Results and discussion**

At low AC current amplitudes $I_0$, the first harmonic dominates in the frequency spectrum of the pick-up coil voltage. If the current amplitude exceeds some threshold value, the amplitude of the first harmonic drops drastically, and the second harmonic becomes dominant. At sufficiently high $I_0$ the frequency spectrum consists of only even harmonics. The even harmonic amplitudes are symmetric with respect to the longitudinal magnetic field and are nonhysteretic. Typical dependencies of several even harmonic amplitudes $V_k$ (where $k$ is the harmonic number) on the longitudinal DC magnetic field $H_e$ at fixed $I_0$ are shown in Fig. 1 for two different wires. The even harmonic amplitudes increase with the DC field, achieve a maximum and then slowly decay. The value of the DC field, at which the harmonic amplitude has a maximum, decreases with the increase of the harmonic number and depends significantly on the AC current amplitude. Fig. 2 shows the dependence of the second harmonic amplitude $V_2$ on $H_e$ for different $I_0$. It follows from Fig. 2 that the value of $V_2$ increases with the AC current amplitude $I_0$ and the permalloy shell thickness.

To interpret the effects observed we use a simple model which allows one to describe main features of the measurements. The AC current $I = I_0 \sin(2\pi f t)$ flowing through the composite wire induces the circular AC magnetic field $H_\varphi$. Simple estimates show that the values of the skin depth in the core and in the shell are larger than the core radius and the shell thickness at 1 MHz for the studied wires. Hence, the skin effect can be neglected, and the circular AC magnetic field distribution within the outer shell can be expressed as

$$H_\varphi(r,t) = (2/cr)I_0 \sin(2\pi ft)\frac{(\sigma_m - \sigma_f)d^2 + 4\sigma_f r^2}{\sigma_m d^2 + \sigma_f(D^2 - d^2)}, \qquad d/2 \leq r \leq D/2 \ . \qquad (1)$$

Here $r$ is the radial coordinate, $\sigma_m$ and $\sigma_f$ are the conductivities of the core and shell, respectively. The circular magnetic field causes a variation of the circular magnetization $M_\varphi$ and, as a result, a change in the longitudinal magnetization $M_z$. In accordance with Faraday law, the variation of $M_z$ induces a voltage $V$ in the pick-up. At low AC current amplitudes the circular field causes a precession of the magnetization vector, and $V$ is proportional to the nondiagonal component of the wire impedance [11]. At higher AC current amplitudes the magnetization reversal process develops, and much higher values of $V$ are induced in the coil.

Since the AC current frequency is comparatively low, the magnetization reversal can be described in the framework of the quasi-static Stoner–Wohlfarth approximation [10]. For simplicity, we neglect the domain structure in the permalloy shell. The minimization of the magnetic free energy consisting of the magnetic anisotropy energy and the Zeeman energy in the fields $H_e$ and $H_\varphi$ yields the following equation for the circular magnetization:

$$(1 - M_\varphi^2)[M_\varphi H_a + M H_\varphi(r,t)]^2 = H_e^2 M_\varphi^2 M^2 \ . \qquad (2)$$

The dependence $M_\varphi(H_\varphi)$ is reversible at $H_e < H_a$ and low $H_\varphi$. If $H_\varphi$ exceeds some threshold value, this dependence acquires the form of the hysteresis loop with the Barkhausen jumps. The threshold current amplitude $I_c$, at which the magnetization reversal develops, calculated from Eqs. (1) and (2) is given by

$$I_c = (cdH_a/4)[1 + \sigma_f(D^2 - d^2)/\sigma_m d^2][1 - (H_e/H_a)^{2/3}]^{3/2} \ . \qquad (3)$$

The characteristic time of the Barkhausen jumps is of the order of $10^{-9}$ s, and the contribution from the jumps to the harmonic amplitudes is insignificant. For the pick-up coil voltage $V$ we have [8]

$$V = \frac{32\pi^3 N f}{c^2} I_0 \cos(2\pi ft) \int_{d/2}^{D/2} \frac{M_\varphi^2 (M^2 - M_\varphi^2) dr}{H_e M_\varphi^3 + H_\varphi(M^2 - M_\varphi^2)^{3/2}} \ , \qquad (4)$$

where $M_\varphi = M_\varphi(r,t)$ is the solution of Eq. (2). The Fourier transformation of Eq. (4) can be used for the analysis of the voltage frequency spectrum. Shown in Fig. 1 is a comparison of the measured and calculated dependencies of the even harmonic amplitudes on the DC field.

It follows from Fig. 1 that the calculated dependencies are in a satisfactory agreement with experimental data both for Cu/NiFe and Nb/NiFe wires for reasonable values of the saturation magnetization $M$ and the anisotropy field $H_a$, which is used as adjusting parameters.

Thus, the observed nonlinear magneto-impedance effect is due to the voltage generated by the magnetization reversal of the wire. The even harmonics dominate in the frequency spectrum of the pick-up coil voltage. The even harmonic amplitudes are sensitive to the longitudinal magnetic field $H_e$. For example, the achieved sensitivity of the second harmonic amplitude is of the order of $50\,\text{mV/Oe}$ at $f = 500\,\text{kHz}$. This sensitivity is sufficiently high and may be enhanced by increasing of the current frequency and amplitude or using the wires having more thick ferromagnetic shell and lower anisotropy field. Hence, the nonlinear magneto-impedance is very promising for the design of weak magnetic field sensors. In conclusion, it should be noted that the cold-drawn composite wires with thick ferromagnetic shell are preferable over the electroplated wires since the pick-up coil voltage is proportional to the volume of the ferromagnetic shell.

**Acknowledgement**

This work was supported in part by the Russian Foundation for Basic Research (Grants 00–02–16707, 01–02–06324 and 01–02–06450). N.A. Buznikov would like to acknowledge the Science Support Foundation, grant for talented young researchers.

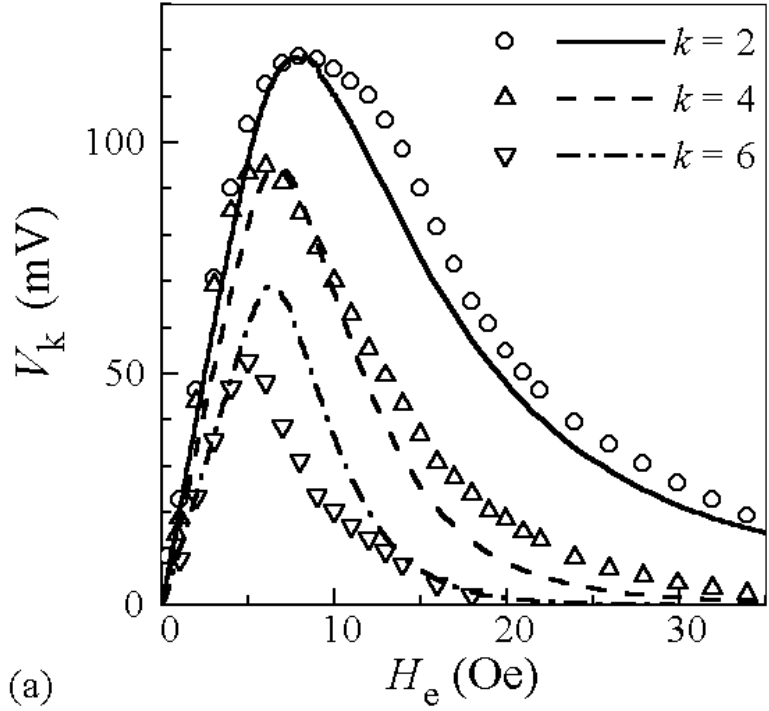

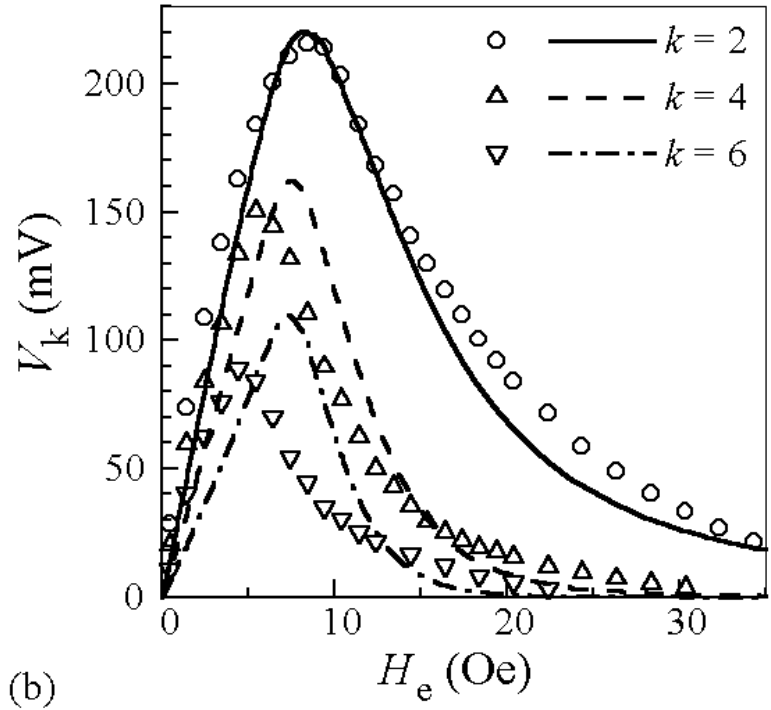

Fig. 1. The dependencies of $V_k$ on $H_e$ at $I_0 = 155\,\text{mA}$ and $f = 500\,\text{kHz}$. (a) Symbols, experimental data for Cu/NiFe composite wire ($d = 50\,\mu\text{m}$, $D = 74\,\mu\text{m}$); lines, calculations at $H_a = 5.8\,\text{Oe}$, $M = 790\,\text{G}$ and $\sigma_f / \sigma_m = 0.02$. (b) Symbols, experimental data for Nb/NiFe composite wire ($d = 65\,\mu\text{m}$, $D = 100\,\mu\text{m}$); lines, calculations at $H_a = 7\,\text{Oe}$, $M = 820\,\text{G}$ and $\sigma_f / \sigma_m = 0.17$.

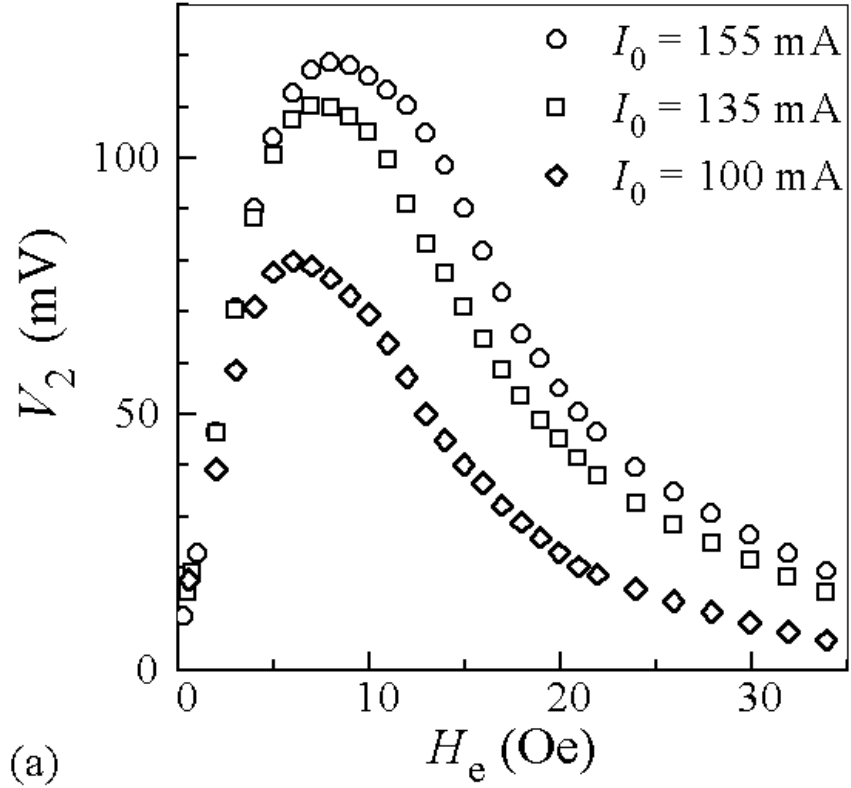

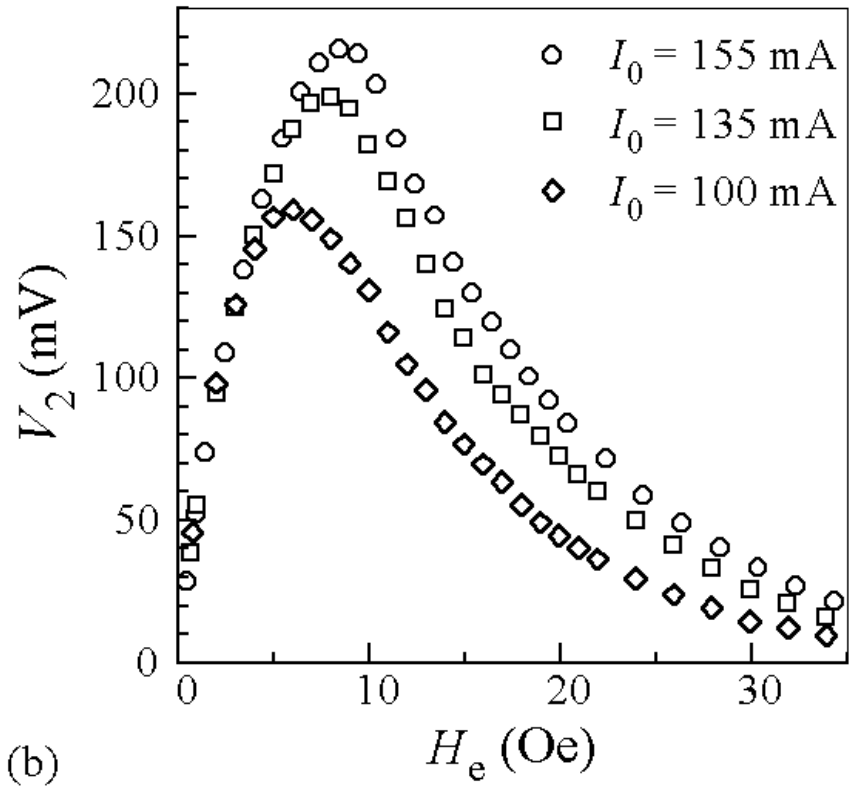

Fig. 2. The dependencies of $V_2$ on $H_e$ at different $I_0$ for (a) Cu/NiFe wire and (b) Nb/NiFe wire. The parameters of wires are the same as in Fig. 1.